\documentstyle[12pt]{article}
\def\fun#1#2{\lower3.6pt\vbox{\baselineskip0pt\lineskip.9pt
\ialign{$\mathsurround=0pt#1\hfil##\hfil$\crcr#2\crcr\sim\crcr}}}

\newcommand{\be}{\begin{equation}}
\newcommand{\ee}{\end{equation}}
\newcommand{\bd}{\begin{displaymath}}
\newcommand{\ed}{\end{displaymath}}
\newcommand{\ba}{\begin{array}}
\newcommand{\ea}{\end{array}}
\newcommand{\bt}{\begin{tabular}}
\newcommand{\et}{\end{tabular}}
\newcommand{\bc}{\begin{center}}
\newcommand{\ec}{\end{center}}

\begin{document}


\hfill {\large\bf YERPHI--1435(5)--95}

\hfill {\large\bf February 1995}
\vspace{1.5cm}

\begin{center}
{\huge\bf Charmed particle production\\[0.4cm]
in hadron--hadron collisions}\\[1.5cm]
G.H.Arakelyan$^1$, P.E.Volkovitsky$^2$\\[0.8cm]

$^1$ Theoretical Department, Yerevan Physics Institute,
Br.Alikhanyan's str. 2,\\ Yerevan 375036, Armenia \\[0.4cm]

$^2$ Institute for Theoretical and Experimental Physics, Moscow
117259, Russia

\end{center}


\vspace{1.5cm}

\begin{abstract}

In the framework of Quark--Gluon--String Model developed recently in
ITEP we calculate spectra of charmed particles $D$, $D_s$,
$\Lambda_c$, $\Xi_c$, $\Omega_c$ in hadron--hadron collisions taking
into account the decays of $S$--wave resonances like $D^*$, $D^*_s$,
$\Sigma_c$, $\Sigma^*_c$, $\Xi^*_c$, $\Xi'_c$, and $\Omega^*_c$.  We
describe the bulk of the existing data on $D$, $D^*$, and $\Lambda_c$
production in $\pi p$ and $pp$ collisions and predict the yield of
charmed particles in $\Sigma^-p$ and $\Xi^-p$ reactions at
hyperon beam energies of $340\;GeV/c$ and $600\;GeV/c$. Because of
significant production of baryon resonances our predictions for
unfavored fragmentation differ from predictions of other models which
do not take resonance production into account.
\end{abstract}

\vfill

\noindent
\underline{\hspace{6.5cm}}

{\small $^1$ E--mail: ARGEV@VXC.YERPHI.AM~~~~~ARGEV@VXITEP.ITEP.RU}

{\small $^2$ E--mail: PETERV@VXITEP.ITEP.RU}

\newpage

\section{Introduction}

{}~~~~~ The physics of charmed particles, e.g. particles which occupy
the place in between light hadrons (hadrons containing $u$--,
$d$-- and $s$--quarks) and heavy hadrons (with $b$--quark), is a
subject of a great interest. On one hand, the methods avaliable for
light hadron sector: quark models, symmetries and so on, can be
applied to charmed hadrons. On the other hand, the presence of
$c$--quark with mass of $1.5-1.8\;GeV$ inside charmed hadrons allows
one to use QCD methods which take this large parameter into account,
like, for example, Effective Heavy Quark Theory (EHQT).

At the moment the spectra of charmed hadrons, both mesons, and baryons
are known much worse than the spectra of light hadrons. One of the
reasons of this situation is due to a small lifetime of stable
charmed particles. The development of new experimental methods
based on the use of vertex detectors gives a hope to have a
significant progress in study of charmed particle spectrum in the
nearest future.

The small lifetime of charmed particles makes it reasonable to look
for charmed particles in experiments with their production in beams
of high energy hadrons, when the lifetime of charmed particles in
laboratory frame increases a few times. One of the goals of the new
generation experiments with $\Sigma^-$ beams to be performed at CERN
(WA89) and at Fermilab (E791) is the study of charmed meson and
baryon spectra.

There is a wide variety of theoretical models which are more or less
satisfactorily applied to description of charmed particle
hadroproduction. These models can be selected into three main groups:
the models based on perturbative QCD and parton model
\cite{1}--\cite{3}, the models of Monte--Carlo type based on Lund
string model \cite{4} and their subsequent development in the form of
event generator PITHIA \cite{5}, and the models which incorporate the
Dual Topological Unitarization scheme \cite{6}--\cite{10}. Among the
last group we will distinguish the model of analytical calculation of
spectra, so called Quark--Gluon String Model (QGSM) developed in ITEP
\cite{11}--\cite{15}.

The use of perturbative QCD based models for calculation of charmed
particle spectra leads to systematic underestimation of these spectra
in the fragmentation region ($x>0.5$). Different variants of
Monte--Carlo models which produce a reasonable description of
$\Lambda_c$ spectra at intermediate energies fail when applied at
higher energies (see, for instance, \cite{16}).

The QGSM was successfully used for description of many features of
multiparticle production, including inclusive spectra of secondary
particles in hadron--hadron collisions both for light
\cite{11}--\cite{15} and charmed \cite{16}--\cite{22} hadrons. In
these papers the production of only stable particles ($\pi$, $K$,
$p$, $\bar p$, $\Lambda$, $\Lambda_c\ldots$) was considered. Spectra
of meson resonances ($\rho$, $f_2$, $K^*$) in the framework of QGSM
were analised in papers \cite{23,24}, but the impact of resonances
into $\pi$ meson spectra was not considered in these papers and the
contributions of direct production and of resonance decay were not
separated.

The main purpose of present paper is to apply the QGSM to
calculation of charmed particle spectra produced in the beams of
negatively charged hyperons, taking into account the decays of
$S$--wave resonances, like $1^-$ mesons ($D^*$ and $D^*_s$),
$1/2^+$ ($\Sigma_c$ and $\Xi'_c$) and $3/2^+$ ($\Sigma^*_c$, $\Xi^*_c$
and $\Omega^*_c$) hyperons. The model parameters we use are obtained
from the description of existing data on charmed particle production.
The contribution of resonances into stable particle spectra will be
taken into account in correspondence with their partial decay widths
\cite{25}.

\section{Model description}

{}~~~~~In the next section we will reanalize in the framework of
the QGSM the inclusive spectra of stable charmed particles
($\Lambda_c$, $\Xi_c$, $\Omega_c$, $D$, $D_s$) taking into account
the contributions from decays of corresponding resonances.

We will calculate only the spectra integrated over transverse momenta
$p_{\bot}$ of produced particles, although there exist
modifications of the QGSM describing $p_{\bot}$ dependence
\cite{22,26}.  $S$--wave charmed resonances decay into stable charmed
particles with emitting $\pi$--meson or $\gamma$--quantum \cite{25},
and we describe the kinematics of this decay as in \cite{27}.

Under such assumptions the invariant cross section of production of
hadron $h$ has the form

\be
\label{1}
x\frac{d\sigma^h}{dx}=x\frac{d\sigma^{h{dir}}}{dx}+
\int\limits^{x^*_+}_{x^*_-}x_R\frac{d\sigma^R}{dx_R}\Phi(x_R)dx_R~.
\ee
Here, $x\frac{d\sigma^{h{dir}}}{dx}$ is the direct production
cross section of hadron $h$, and $x_R\frac{d\sigma^R}{dx_R}$ is
the $R$--resonance production cross section. Function $\Phi(x_R)$
describes two--body decays of resonance $R$ into hadron $h$. After
integration over transverse momenta of both hadron $h$ and resonance
$R$ the function $\Phi(x_R)$ has the form

\be
\label{2}
\Phi(x_R)=\frac{M_R}{2p^*}\frac{1}{x^2_R}~.
\ee
In eqs. (\ref{1}) and (\ref{2}) $x_R$ is the Feynman variable
of resonance $R$

\bd
x^*_+=\frac{M_R\tilde x}{E^*-p^*},\quad x^*_-=\frac{M_R\tilde
x}{E^*+p^*},\quad \tilde x=\sqrt{x^2+x^2_{\bot}},\quad
x_{\bot}=\frac{2\sqrt{<p^2_{\bot}>+m^2}}{\sqrt s},
\ed
$m$ is the mass of produced hadron $h$, $M_R$ is the mass of
resonance, $E^*$ and $p^*$ are energy and $3$--momentum of hadron $h$
in the resonance rest frame, $<p^2_{\bot}>$ is the average transverse
momentum squared of hadron $h$.

The inclusive spectra of hadron $h$ in the framework of the QGSM has
the form \cite{11}--\cite{15}

\be
\label{3}
x\frac{d\sigma^h}{dx}=\sum^{\infty}_{n=0}\sigma_n(s)\varphi^h_n(x)~,
\ee
where $\sigma_n(x)$ is the cross section of $n$--pomeron shower
production and $\varphi^h_n(x)$ determines the contribution of
diagram with $n$ cut pomerons.

The expressions for $\sigma_n(s)$ and corresponding parameter values
for $pp$ and $\pi p$ collisions are given in \cite{11}--\cite{15}.
For $\Sigma^-p$ interaction author of paper \cite{20} used the
same values as for $pp$. In the framework of additive quark model we
calculate the relation between pomeron residues in
$\Sigma^-p(\Xi^-p)$ and $pp$ collisions

\bd
\gamma_{\Sigma p}=0.92\gamma_{pp};\quad \gamma_{\Xi
p}=0.84\gamma_{pp}
\ed
The function $\varphi^h_n(x)$ $(n>1)$ for $\pi p$ interaction can be
written in the form \cite{11}--\cite{15}

\be
\label{4}
\varphi^h_n(x)=f^h_{\bar
q}(x_+,n)f^h_q(x_-,n)+f^h_q(x_+,n)f^h_{qq}(x_-,n)+2(n-1)f^h_s(x_+,n)
f^h_s(x_-,n)
\ee
and for baryon--proton interaction

\be
\label{5}
\varphi^h_n(x)=f^h_{qq}(x_+,n)f^h_q(x_-,n)+f^h_q(x_+,n)f^h_{qq}(x_-,n)+
2(n-1)f^h_{sea}(x_+,n)f^h_{sea}(x_-,n)~,
\ee
where $x_{\pm}=\frac{1}{2}[\sqrt{x^2+x^2_{\bot}}\pm x]$.

The functions $f^h_i(x,n)(i=q,\bar q,qq,q_{sea})$ in (\ref{4}) and
(\ref{5}) describe the contributions of the valence/sea quarks,
antiquarks and diquarks, respectively. They represent a convolution of
quark/diquark momentum distribution functions $u_i(x,n)$ in the
colliding hadrons and of  the function of
quark/diquark fragmentation into hadron "$h$" $G^h_i(x,n)$:

\be
\label{6}
f_i(x,n)=\int^1_x u_i(x_1,n)G_i(x/x_1)dx_1~.
\ee
The projectile (target) contribution depends on the valiable $x_+$
$(x_-)$.

The quark functions $f_q(x,n)$ in the case of $\pi p$ and
$pp$--collisions are given in \cite{11}--\cite{15} without taking
into account the possible resonance production.

For the $\Sigma^-$ $(\Xi^-)$ beams the functions $f^h_q(x,n)$ are
expressed in terms of corresponding $s$-- ($f^h_s(x,n)$) and $d$--
($f^h_d(x,n)$) quark functions in the following form

\be
\label{7}
\ba{ll}
f^{h(\Sigma^-)}_q(x,n)=\frac13f^{h(\Sigma^-)}_s(x,n)+
\frac23f^{h(\Sigma^-)}_d(x,n)\\[3mm]
f^{h(\Xi^-)}_q(x,n)=\frac23f^{h(\Xi^-)}_s(x,n)+
\frac13f^{h(\Xi^-)}_d(x,n)~.
\ea
\ee

In the framework of the additive quark model diquarks in $S$--wave
baryons may have spin (isospin) $0$ and $1$. So the diquark functions
$f^h_{qq}(x)$ are expressed in terms of scalar $(0)$ and vector $(1)$
diquark functions with the weights determined by $SU(6)$ symmetric
functions \cite{28}.

\be
\label{8}
\ba{lll}
f^{h(p)}_{qq}=\frac13f^{h(p)}_{uu}(x,n)+\frac12f^{h(p)}_{(ud)_0}(x,n)+
\frac16f^{h(p)}_{(ud)_1}(x,n)\\[3mm]
f^{h(\Sigma^-)}_{qq}=\frac13f^{h(\Sigma^-)}_{dd}(x,n)+
\frac12f^{h(\Sigma^-)}_{(ds)_0}(x,n)+
\frac16f^{h(\Sigma^-)}_{(ds)_1}(x,n)\\[3mm]
f^{h(\Xi^-)}_{qq}=\frac13f^{h(\Xi^-)}_{ss}(x,n)+
\frac12f^{h(\Xi^-)}_{(ds)_0}(x,n)+
\frac16f^{h(\Xi^-)}_{(ds)_1}(x,n)~.
\ea
\ee

In what follows, we will assume that distribution functions of scalar
and vector diquarks $u_{qq}(x,n)$ are the same. Of course, different
diquarks fragment into baryons in different ways. So, for
instance, the direct production of $\Lambda_c$ in $pp$ collision is
determined by scalar (and isoscalar) diquark function $f_{(ud)_0}$
(Fig.1a), and direct production of $\Sigma_c$ and $\Sigma^*_c$
hyperons are determined by the vector diquark function $f_{(ud)_1}$.

We also assumed that spin of diquark do not influence
splitting of diquark (one of such terms are shown on Fig.1b). In the
case of diquark fragmentation into meson this assumption leads to
equation $f^M_{(qq)_0}(x,n)=f^M_{(qq)_1}(x,n)$ and eq.(\ref{8})
reduces to the expressions

\be
\label{9}
\ba{lll}
f^{M(p)}_{qq}(x,n)=\frac13f^{M(p)}_{uu}(x,n)+
\frac23f^{M(p)}_{ud}(x,n)\\[3mm]
f^{M(\Sigma^-)}_{qq}(x,n)=\frac13f^{M(\Sigma^-)}_{dd}(x,n)+
\frac23f^{M(\Sigma^-)}_{us}(x,n)\\[3mm]
f^{M(\Xi^-)}_{qq}(x,n)=\frac13f^{M(\Xi^-)}_{ss}(x,n)+
\frac23f^{M(\Xi^-)}_{ds}(x,n)
\ea
\ee
which coincide with \cite{20} for the diquark in proton and
$\Sigma^-$--hyperon.

A full list of the quark/diquark distribution functions in
$\pi$--meson, $p$, $\Sigma^-$, and $\Xi^-$--hyperons used in this
work is given in Appendix I.

Further, we will assume that fragmentation functions of the quarks
and diquarks does not depend on the spin of the picked up quark (or
diquark). From this assumption the equality of the fragmentation
functions of the corresponding quarks or diquarks to $\Sigma_c$-- and
$\Sigma^*_c$--, $\Xi'_c$-- and $\Xi^*_c$--baryons, $D$-- and
$D^*$--mesons follows. The method developed in the present paper for
parametrization of the fragmentation function slightly differs from
the one used in \cite{11}--\cite{24}. We represent the fragmentation
function as a sum of two terms. The first one is parametrized as a
product of two polynomials each of which corresponds to the sum of
all possible assymptotic and preasymptotic terms in the fragmentation
($x\to1$, expanding in the series of $(1-x)$) and central ($x\to0$,
expanding in the series of $x$) regions. The second term stands for
the case when the fragmented object neither as a whole nor even
partly gets into created hadron. In the case of nonleading
fragmentation the corresponding functions are parametrized only using
the second term.  Appendix II contains the full list of fragmentation
functions of quarks and diquarks into $\Lambda_c$, $\Sigma_c$,
$\Sigma^-_c$, $\Xi_c$, $\Xi'_c$, $\Xi^*_c$, $\Omega_c$,
$\Omega^*_c$--baryons and $D$, $D^*$, $D_s$, $D^*_s$--mesons. The
values of free parameters, determined from the comparison with the
experimental data on $\Lambda_c$ production in $\pi p$ \cite{29} and
$pp$ \cite{30,31} collitions, $D$ and $D^*$--mesons in $pp$
\cite{33}--\cite{35} and $\pi p$ \cite{35}--\cite{38} interation are
also given in Appendix II.

\section{Comparison with experimental data and the predictions of
the model}

\subsection{Description of charmed hadron spectra}

{}~~~~~In this section, we consider the description of the existing
experimental data for the $\Lambda_c$--hyperon and $D$ and
$D^*$--meson production in $\pi p$ and $pp$--collisions in the
framework of the present model.

The $\Lambda_c$--baryon spectra in $\pi^-p$--collision
at $230GeV/c$ \cite{24} are shown in Fig.2a,b correspondingly. The
theoretical curves are  calculated by using eq.(\ref{1}) taking
into account the contributions of $\Lambda_c$ produced in decays
$\Sigma_c\to\Lambda_c\pi$ and $\Sigma^*_c\to\Lambda_c\pi$. The
dotted line in Fig.2b shows the contribution of direct
$\Lambda_c$. As one can see the agreement with $\pi^-p$ data is
satisfactory. Concerning the data \cite{30,31} in $pp$--collision one
can see a noticeable difference between the data of groups \cite{30}
and \cite{31}. This uncertainty does not allow us to have the
unambiguous values of model parameters. The values given in
Appendix II correspond to the curve plotted in Fig.2. Our
calculations of $\Lambda_c$ integral cross sections in $\pi p$ and
$pp$ interactions are compared with corresponding experimental data
in Table I. Large uncertainity in $\Lambda_c$ integral cross section
presented in paper \cite{30} are mainly due to method
used for extrapolation of the data to low $x$ region.

The inclusive $x_F$ distribution of $D$--mesons produced in $pp$ and
$\pi p$ collision \cite{33}--\cite{38} are compared with our
calculations in Figs.3--5. Both theoretical predictions for the
integral cross sections and experimental data are shown in Table 2.

In Figs.3a--d we plotted the experimental points for the spectra of
different $D$--mesons produced in $pp$--collision at $400\;GeV/c$
\cite{32} together with QGSM calculations.

Inclusive distributions of all $D$--mesons in $pp$ interaction at
momenta of $200\;GeV$ \cite{35}, $400\;GeV/c$ \cite{32} and
$800\;GeV/c$ \cite{34} are presented in Fig.4a--c. The theoretical
curves are calculated for the sums of spectra of all $D$--mesons.

Fig.5a--d present a comparison of the $x_c$--distribution of leading
($D^-$ and $D^0$) and non--leading ($D^+$ and $\bar D^0$) charmed
mesons in $\pi^-p$ interaction at $200\;GeV/c$ \cite{35} and
$360\;GeV/c$ \cite{37} with theoretical calculations.

As far as we  consider here for the first time the
QGSM spectra of resonances taking into account their
subsequent decays, it is important to compare our calculations with
the available data on $D^*$--meson productions in $pp$ \cite{32} and
$\pi p$ \cite{38} collitions. The data on reactions $\pi^-p\to
D^{*+}/D^{*-}X$ and $\pi^-p\to D^{*0}/\bar D^{*0}X$ at $360\;GeV/c$
\cite{38} are compared with our predictions in Fig.6a,b. It seems
that the agreement for the sum of spectra of $D^{*+}$ and $D^{*-}$
mesons is rather reasonable. As to the neutral mesons $D^{*0}$ and
$\bar D^{*0}$, there the experimental information is rather scarce.
The curves in Fig.6 correspond to the sums of spectra of $D^{*+}$ and
$D^{*-}$ (Fig.6a) and $D^{*0}$ and $\bar D^{*0}$ (Fig.6b) mesons.
Table III contains the experimental data on integral cross sections
on $D^*$ meson production in $pp$ and $\pi p$ collision together
with our calculations. The upper limit on $D^{\pm}_s$--meson
integral cross section in $pp$--interaction \cite{32} is
also shown in Table III together with our calculations. As one can
see, agreement of model calculations for above mentioned inclusive
spectra is rather good.

\subsection{Model predictions for charmed hyperon production with
hyperon beam}

{}~~~~~In this section we consider the predictions for different
inclusive spectra of charmed hadrons on $\Sigma^-$ and $\Xi^-$ beams.
We restrict ourselves to initial momenta of $340$ and $600\;GeV/c$
which correspond to the existing experiments in CERN (WA89) and FNAL
(E791). The parameters used were obtained from description of
experimental data on charmed hadrons in $pp$ and $\pi p$ collisions.
The inclusive spectra of stable baryon states $\Lambda_c$,
$\Xi^+_c$, $\Xi^0_c$ and $\Omega_c$ at $340\;GeV/c$ are shown in
Fig.7a and  the spectra of $\Sigma^{*0}_c$, $\Xi^{*0}_c$,
$\Xi^{*+}_c$ and $\Omega^{*0}_c$ are plotted  in Fig.7b. The same
spectra for $600\;GeV/c$ are given in Fig.7c,d.

Fig.8a--d present our calculations for the same particles (except
$\Lambda_c$) obtained in $\Xi^-p$ collision. The predictions for
Feynman--x distributions of spectra of $D$ and $D^*$ mesons are shown
in Fig.9a,b ($340\;GeV/c$) and Fig.9c,d ($600\;GeV/c$). The same
calculations for $\Xi^-$ beam are presented in Fig.10a--d.

It is interesting to consider the predictions for $D_s$ and $D^*_s$
meson production in hyperon beams. As far as $D^-_s$ and $D^{*-}_s$
contain valence $s$--quark from initial hyperon beam, the
cross-section of these mesons should not be small. Our calculations
for the same momenta are pictured in Fig.11a--d
($\Sigma^-$--beam) and Fig.12a--d ($\Xi^-$--beam).

\section{Conclusion}

{}~~~~~In our paper a modification of QGSM which takes into
account the resonance decay contributions is presented. The main
result of our analysis of experimental data can be summarized as
following:

--- The model under consideration describes with a good accuracy the
inclusive $x$--spectra of charmed particles in $\pi p$-- and
$pp$--collisions. So we hope that our predictions for charmed
particle production of hyperon beams are fairly reliable.

--- The model predicts fairly large yield of resonances and as a
result values of the spectra of unfavored stable particles in the
fragmentation region are rather close to the spectra of
favored particles. As a striking example we note the
$\Sigma^-p\to\Xi^+_cX$ reaction where at $x_F>0.2$ the $\Xi^+_c$
spectra are completely determined by leading produced
$\Xi^{*0}_c$--resonances via $\Xi^{*0}_c\to\Xi^+_c\pi^-$ decay.

Let us note that it is possible to generalise the present model
and include the dependence on transverse momenta of produced particle
taking into account the intrinsic charm and also to consider processes
of production of $P$--wave resonances. In our opinion
we have well described the data in the fragmentation region
($x_F>0.2$). More accurate description of the experimental
data requires a more careful consideration of the resonance decay
kinematics, that is essentially connected with taking into
account the transverse momentum dependence.  \bigskip

{\it Acknowlegments}. We are grateful to A.B.Kaidalov,
K.G.Boreskov, M.A.Kubantsev, O.I.Piskounova, and  for useful
discussion.  G.H.Arakelyan would like to thank the Theoretical
Department of ITEP for kind hospitality during his visits to ITEP
where this work was done. This work was supported in part by
International Science Foundation Grant $N$ RYE000 and Grant
INTAS--93--79.

\newpage

\section*{Appendix I. The distribution functions of quarks and
diquarks in the projectile and target hadrons}

\setcounter{equation}{0}
\def\theequation{A.\arabic{equation}}

{}~~~~~The distribution functions of quarks (diquarks) in hadron $h$
are parametrized in the standard form

\be
\label{a1.1}
f^n_i(x,n)=C_ix^{\alpha_i}(1-x)^{\beta'_i}~,
\ee
where $\beta'_i=\beta_i+2(n-1)(1-\alpha^0_{\rho})$. The coefficients
$C_i$ in (\ref{a1.1}) are determined by the normalization condition
$\int^1_0f^n_i(x,n)dx=1$ and are equal to

\be
\label{a1.2}
C_i=\frac{\Gamma(1+\alpha)\Gamma(1+\beta')}{\Gamma(2+\alpha+\beta)}~.
\ee
Here, $\Gamma(\alpha)$ is the Gamma--function.

The values of $\alpha$ and $\beta$ can be expressed in terms of Regge
trajectory intercepts and are shown in Table A.1.
We use the values of $\alpha^0_{\rho}=0.5$, $\alpha^0_{\varphi}=0$,
$\alpha^0_N=-0.5$.

\section*{Appendix II. The functions of quark and diquark
fragmentation into charmed hadrons}

{}~~~~~The fragmentation functions for both quarks and diquarks consist
of two terms. The first one appears when the object $i$ under
consideration (quark, diquark as a whole or some its part) enters
into the produced hadron $h$ (favored fragmentation, see Fig.1a). The
second term corresponds to the absence of object $i$ in the hadron
$h$ (unfavored fragmentation).

\be
\label{a2.1}
G^h_i(x)=G^h_{1i}(x)+G^h_{2i}(x)
\ee

Functions $G^n_{1i}(x)$ and $G^h_{2i}(x)$ in the $x\to1$ region
contain the universal factor

\be
\label{a2.2}
F_1(x)=(1-x)^{\lambda-\alpha^0_{\psi}}
\ee

For convenience we will further consider the fragmentation functions
of quark and diquarks separately.

\subsection*{II.1 Quark fragmentation functions}

{}~~~~~In the case of leading fragmentation functions $G^h_{1i}(x)$ are
parametrized in the form

\be
\label{a2.3}
G^h_{1i}(x)=d^hx^{\varepsilon_i}F_1(x)(1-x)^{\gamma_k}
\ee
where $i=q(u,d),s$ is the type of the fragmented quark, $k$ is the
type of object (quark, antiquark or diquark) which combines
with quark $i$ to produce hadron $h$. The expressions for
$\varepsilon_i$ (lines 1, 2 for the case of quark fragmentation) and
$\gamma_k$ are given in Tables A.2 \footnote{Table A.2 contains also
the values for diquark fragmentation.} and A.3. Let us note that
$\gamma_k$ does not depend on the type of fragmented quark.

The second term in (\ref{a2.1}) is parametrized in the form

\be
\label{a2.4}
G^h_{2i}(x)=a^h_0F_1(x)(1-x)^{\delta_m}~.
\ee
Here, $m$ depends on the quark content of produced hadron $h$ and the
type of fragmented quark.

In the case of nonleading fragmentation function $G^h_i(x)$
contains only the second term given by eq.(\ref{a2.4}). The $\delta_m$
is expressed in terms of intersepts of Regge trajectories as
presented in Table A.4 (lines 1--3).

\subsection*{II.2 Diquark fragmentation functions}

{}~~~~~Since only one quark from diquark takes part in meson
creation, diquark fragmentation function into mesons is described by
the same equations as the quark fragmentation function into baryon:
eqs.(\ref{a2.1}--\ref{a2.4}) with the same values of $\varepsilon_l$,
$\gamma_k$ and $\delta_m$. In this case $l=q,s$ are the quarks from
diquarks which take part into meson creation, $k=qjjc$, $sjjc$
($j$ denotes string junction) is the part of diquark width does not
enter meson $h$ and $c$--quark with string junction (see lines 3
and 4 in Table A.3). $\delta_m$, where $m$ stands for sum of all
components of fragmented and created objects, is given in lines 1--3
of Table A.4. If only one of the quarks from diquark enters into
produced meson, for instance $G^{D^-}_{ds}(x)$, $G^{D^-_s}_{ds}(x)$,
$G^{\bar D^0}_{ud}(x)$ and $G^{D^-}_{ud}(x)$, eq.(\ref{a2.2}) is
divided by factor 2.

As to diquark fragmentation into baryons, the functions $G^h_{1i}(x)$
contain both leading term, described by diagram in Fig.1a and terms,
corresponding to the splitting of diquark (one of them is shown in
Fig.1b). In this case $G^h_{1i}(x)$ is parametrized as

\be
\label{a2.5}
G^h_{1i}(x)=(\sum_l b_lx^{\varepsilon_l})\cdot(a_1+\sum_{k>1}
a_k(1-x)^{\gamma_k})F_1(x)~.
\ee
If fragmentation functions do not contain the leading term,
$G^h_{1i}(x)$ has the form

\be
\label{a2.6}
G^h_{1i}(x)=(\sum_l b_lx^{\varepsilon_l})\cdot(\sum_{k>1}
a_k(1-x)^{\gamma_k})F_1(x)~.
\ee

In eqs.(\ref{a2.5}) and (\ref{a2.6}) $l$ means the type of objects
from diquark $i$ which take part in creation of baryon $h$ (except
the whole diquark) in $x\to0$ region. The values of
$\varepsilon_l$ are taken from Table A.2. $k$ in (\ref{a2.5}) and
(\ref{a2.6}) is the type of diquark splitting diagram (one of them
is shown in Fig.1b) and denotes the remained part of diquark
which do not form baryon $h$ and the objects combined baryon $h$
together with objects $l$.  Table A.5 contains the values of
$\gamma_k$ expressed in terms of of Regge trajectory intercepts
for charmed baryon production.

If only one quark or quark with string junction from fragmented
diquark $i$ passed into the created baryon, the corresponding
constants $b_l$ and $a_k$ are divided by factor 2.

The function $G^h_{2i}(x)$ for diquark fragmentation into baryon
is also parametrized in the form (\ref{a2.4}). The corresponding
values of $\delta_m$ are given in Table A.6. Here $m$ is the sum over
all types of objects in fragmented diquark $i$ and created baryon $h$.
The parameter values which were taken from the comparison with
experimental data are equal to

\bd
\ba{lllll}
a^D_0=a^{D^*}_0=a^{\Lambda_c}_0=a^{\Sigma^*_c}_0=a_0=0.023\\
a^{\Xi_c}_0=a^{\Xi'_c}_0=a^{\Xi^*_c}_0=a^{D_s}_0=a^{D^*_s}_0=a_0\delta\\
a^{\Omega_c}_0=a^{\Omega^*_c}_0=a_0\delta^2\\
a_1=4,\quad a_2=a_3=a_4=d=0.23,\quad a_5=\ldots=a_{13}=a_0\\
b_1=b_2=1,\quad b_3=b_4=b_5=0,\quad b_6=b_7=0.1,\quad b_8=0.01~,
\ea
\ed
$\delta=\frac{1}{3}$ stands for $s$--pair supression. Intercepts
$\alpha^0_{\rho}$, $\alpha^0_{\varphi}$, $\alpha^0_N$ are given in
Appendix I. According to paper \cite{39} we used only
$\alpha^0_{\psi}=-2.18$

\newpage

\noindent
\begin{footnotesize}
\bc
{\Large\bf Table A.1}\\[0.3cm]
\bt{|c|c|c|c|c|c|c|c|c|}\hline
\multicolumn{1}{|r|}{$n$} & \multicolumn{2}{c|}{$\pi^-$} &
\multicolumn{2}{c|}{$p$} & \multicolumn{2}{c|}{$\Sigma^-$} &
\multicolumn{2}{c|}{$\Xi^-$} \\ \hline
\multicolumn{1}{|l|}{$i$} & $\alpha$ & $\beta$ & $\alpha$ & $\beta$ &
$\alpha$ & $\beta$ & $\alpha$ & $\beta$ \\ \hline

&&&&&&&& \\

$u$ & & & $-\alpha^0_{\rho}$ & $\alpha^0_{\rho}-2\alpha^0_N$ &
$-\alpha^0_{\rho}$ & $\alpha^0_{\rho}-2\alpha^0_N$ &
$-\alpha^0_{\rho}$ & $\alpha^0_{\rho}-2\alpha^0_N$ \\
& & & & & & $+\alpha^0_{\rho}-\alpha^0_{\varphi}$ & &
$+2(\alpha^0_{\rho}-\alpha^0_{\varphi})$ \\

&&&&&&&& \\
\hline
&&&&&&&& \\

$d$ & $-\alpha^0_{\rho}$ & $-\alpha^0_{\rho}$ & $\alpha^0_{\rho}$ &
$\alpha^0_{\rho}-2\alpha^0_N$ & $-\alpha^0_{\rho}$ &
$\alpha^0_{\rho}-2\alpha^0_N$ & $-\alpha^0_{\rho}$ &
$\alpha^0_{\rho}-2\alpha^0_N$ \\
& & & $-\alpha^0_{\rho}$ & $+1$ & &
$+\alpha^0_{\rho}-\alpha^0_{\varphi}$ & &
$+2(\alpha^0_{\rho}-\alpha^0_{\varphi})$ \\

&&&&&&&& \\
\hline
&&&&&&&& \\

$\bar u$ & $-\alpha^0_{\rho}$ & $-\alpha^0_{\rho}$ & $-\alpha^0_{\rho}$ &
$\alpha^0_{\rho}-2\alpha^0_N$ & $-\alpha^0_{\rho}$ &
$\alpha^0_{\rho}-2\alpha^0_N$ & $-\alpha^0_{\rho}$ &
$\alpha^0_{\rho}-2\alpha^0_N$ \\

& & & & & & $+\alpha^0_{\rho}-\alpha^0_{\varphi}$ & &
$+2(\alpha^0_{\rho}-\alpha_{\varphi})$ \\

&&&&&&&& \\
\hline
&&&&&&&& \\

$\bar d$ & $-\alpha^0_{\rho}$ & $-\alpha^0_{\rho}$ & $-\alpha^0_{\rho}$ &
$\alpha^0_{\rho}-2\alpha^0_N$ & $-\alpha^0_{\rho}$ &
$\alpha^0_{\rho}-2\alpha^0_N$ & $-\alpha^0_{\rho}$ &
$\alpha^0_{\rho}-2\alpha^0_N$ \\
& & & & $+1$ & & $+\alpha^0_{\rho}-\alpha^0_{\varphi}$ & &
$+2(\alpha^0_{\rho}-\alpha^0_{\varphi})$ \\

&&&&&&&& \\
\hline
&&&&&&&& \\

$s$ & $-\alpha^0_{\varphi}$ & $-\alpha^0_{\varphi}$ &
$-\alpha^0_{\varphi}$ & $\alpha^0_{\rho}-2\alpha^0_N$ &
$-\alpha^0_{\varphi}$ & $\alpha^0_{\rho}-2\alpha^0_N$ &
$-\alpha^0_{\varphi}$ & $\alpha^0_{\rho}-2\alpha^0_N$ \\
& & & & $+\alpha^0_{\rho}-\alpha^0_{\varphi}$ & & & &
$+\alpha^0_{\rho}-\alpha^0_{\varphi}$ \\

&&&&&&&& \\
\hline
&&&&&&&& \\

$\bar s$ & $-\alpha^0_{\varphi}$ & $-\alpha^0_{\varphi}$ &
$-\alpha^0_{\varphi}$ & $\alpha^0_{\rho}-2\alpha^0_N$ &
$-\alpha^0_{\varphi}$ & $\alpha^0_{\rho}-2\alpha^0_N$ &
$-\alpha^0_{\rho}$ & $\alpha^0_{\rho}-2\alpha_N$ \\
& & & & $+\alpha^0_{\rho}-\alpha^0_{\varphi}$ & &
$+2(\alpha^0_{\rho}-\alpha^0_{\varphi})$ & &
$+3(\alpha^0_{\rho}-\alpha^0_{\varphi})$ \\

&&&&&&&& \\
\hline
&&&&&&&& \\

$uu$ & & & $\alpha^0_{\rho}-2\alpha^0_N$ & & & & & \\
$dd$ & & & $+1$ & $-\alpha^0_{\rho}$ &
$\alpha^0_{\rho}-2\alpha^0_N$ & $-\alpha^0_{\varphi}$ & & \\
&&&&&&&& \\

$ud$ & & & $\alpha^0_{\rho}-2\alpha^0_N$ & $-\alpha^0_{\rho}$ & & & &
\\

&&&&&&&& \\
\hline
&&&&&&&& \\

$us$ & & & & &
$\alpha^0_{\rho}-2\alpha^0_N$ & $-\alpha^0_{\rho}$ &
$\alpha^0_{\rho}-2\alpha^0_N$ &
$-\alpha^0_{\varphi}$ \\
$ds$ & & & & & $+\alpha^0_{\rho}-\alpha^0_{\varphi}$ & &
$+\alpha^0_{\rho}-\alpha^0_{\varphi}$ & \\
&&&&&&&& \\
\hline
&&&&&&&& \\

$ss$ & & & & & & &
$\alpha^0_{\rho}-2\alpha^0_N$ &
$-\alpha^0_{\rho}$ \\
& & & & & & & $+2(\alpha^0_{\rho}-\alpha^0_{\varphi})$ & \\
&&&&&&&& \\
\hline
\et
\ec
\end{footnotesize}

\newpage

\bc
{\Large\bf Table A.2}\\[0.3cm]
\bt{|c|c|c|}\hline
$N$ & $l$ & $\varepsilon_l$ \\ \hline

$1$ & $q$ & $1-\alpha^0_{\rho}$ \\
&&\\

$2$ & $s$ & $1-\alpha^0_{\varphi}$ \\
&&\\

$3$ & $q+j$ & $2(\alpha_{\rho}-\alpha_N)$ \\
&&\\

$4$ & $s+j$ & $2(\alpha_{\rho}-\alpha_N)+\alpha_{\rho}-\alpha_{\varphi}$ \\
&&\\

$5$ & $qq$ & $2(1-\alpha_{\rho})$ \\
&&\\

$6$ & $qs$ & $2-\alpha_{\rho}-\alpha_{\varphi}$ \\
&&\\

$7$ & $ss$ & $2(1-\alpha_{\varphi})$ \\
&&\\

$8$ & $j$ & $3\alpha_{\rho}-2\alpha_N-1$ \\ \hline
\et
\ec

\bigskip

\bc
{\Large\bf Table A.3}\\[0.3cm]
\bt{|c|c|c|}\hline
$N$ & $k$ & $\gamma_k$ \\ \hline

$1$ & $q$ & $0$ \\
&&\\

$2$ & $s$ & $0$ \\
&&\\

$3$ & $qcj$ & $2(\alpha_{\rho}-\alpha_N)$ \\
&&\\

$4$ & $scj$ & $2(\alpha_{\rho}-\alpha_N)+\alpha_{\rho}-
\alpha_{\varphi}$ \\ \hline
\et
\ec

\bigskip

\bc
{\Large\bf Table A.4}\\[0.3cm]
\bt{|c|c|c|}\hline
$N$ & $m$ & $\delta_m$ \\ \hline

$1$ & $qq\bar c$ & $2(1-\alpha_{\rho})$ \\
&&\\

$2$ & $qs\bar c$ & $2-\alpha_{\rho}-\alpha_{\varphi}$ \\
&&\\

$3$ & $ss\bar c$ & $2(1-\alpha_{\varphi})$ \\
&&\\

$4$ & $3qcj$ & $2(1-\alpha_N)$ \\
&&\\

$5$ & $2qscj$ & $2(1-\alpha_N)+\alpha_{\rho}-\alpha_{\varphi}$ \\
&&\\

$6$ & $q2scj$ & $2(1-\alpha_N)+2(\alpha_{\rho}-\alpha_{\varphi})$ \\
&&\\

$7$ & $3scj$ & $2(1-\alpha_N)+3(\alpha_{\rho}-\alpha_{\varphi})$ \\
\hline
\et
\ec

\newpage

\bc
{\Large\bf Table A.5}\\[0.3cm]
\bt{|c|c|c|}\hline
$N$ & $k$ & $\gamma_k$ \\ \hline

$1$ & $c$ & $0$ \\
&&\\

$2$ & $qjqcj$ & $4(\alpha_{\rho}-\alpha_N)$ \\
&&\\

$3$ & $qjscj$ &
$4(\alpha_{\rho}-\alpha_N)+\alpha_{\rho}-\alpha_{\varphi}$ \\
&&\\

$4$ & $sjscj$ &
$4(\alpha_{\rho}-\alpha_N)+2(\alpha_{\rho}-\alpha_{\varphi})$ \\
&&\\

$5$ & $qqc$ & $2(1-\alpha_{\rho})$ \\
&&\\

$6$ & $qsc$ & $2-\alpha_{\rho}-\alpha_{\varphi}$ \\
&&\\

$7$ & $ssc$ & $2(1-\alpha_{\varphi})$ \\
&&\\

$8$ & $4uc$ & $4(1-\alpha_{\rho})$ \\
&&\\

$9$ & $3qsc$ & $4(1-\alpha_{\rho})+\alpha_{\rho}-\alpha_{\varphi}$ \\
&&\\

$10$ & $2q2sc$ &
$4(1-\alpha_{\rho})+2(\alpha_{\rho}-\alpha_{\varphi})$ \\
&&\\

$11$ & $q3sc$ & $4(1-\alpha_{\rho})+3(\alpha_{\rho}-\alpha_{\phi})$ \\
&&\\

$12$ & $4sc$ & $4(1-\alpha_{\varphi})$ \\
&&\\

$13$ & $jjc$ & $6\alpha_{\rho}-4\alpha_N-2$ \\ \hline
\et
\ec

\bigskip

\bc
{\Large\bf Table A.6}\\[0.3cm]
\bt{|c|c|c|}\hline
$N$ & $m$ & $\delta_m$ \\ \hline

$1$ & $4qc2j$ & $2\alpha_{\rho}-4\alpha_N+2$ \\
&&\\

$2$ & $3qsc2j$ &
$-2\alpha_{\rho}-4\alpha_N+2+\alpha_{\rho}-\alpha_{\varphi}$ \\
&&\\

$3$ & $2q2sc2j$ &
$2\alpha_{\rho}-4\alpha_N+2+2(\alpha_{\rho}-\alpha_{\varphi})$ \\
&&\\

$4$ & $q3sc2j$ &
$2\alpha_{\rho}-4\alpha_N+2+3(\alpha_{\rho}-\alpha_{\varphi})$ \\
&&\\

$5$ & $4sc2j$ &
$2\alpha_{\rho}-4\alpha_N+2+4(\alpha_{\rho}-\alpha_{\varphi})$ \\
\hline
\et
\ec

\newpage

\bc
{\Large\bf Table I}\\[0.3cm]
\bt{|c|c|c|c|c|}\hline
Reaction & Ref. & \parbox{2cm}{$P_L\;(GeV/c)$\\ or $\sqrt s\;(GeV)$} &
$\sigma_{exp}(\mu b)$ & $\sigma_{theor}(\mu b)$\\ \hline
&&&& \\

$pp\to\Lambda_c~X$ & & & $40\pm18$ & \\
& \cite{30} & $63\;GeV$ & $204\pm11$ & $660$ \\
all $x$ & & $$ & $2046\pm836$ & $$ \\
&&&& \\\hline

&&&& \\
$pp\to\Lambda_c~X$ & \cite{31} & $63\;GeV$ & $101\pm18\pm26$ & $84$ \\
$|x|>0.5$ & & & & \\
&&&& \\\hline

&&&& \\
$\pi^-N\to\Lambda_c~X$ & \cite{29} & $230\;GeV/c$ & $4.9\pm1.4\pm0.7$
& $6.8$ \\
$x_c>0$ & & $$ & $$ & $$ \\
&&&& \\\hline
\et
\ec

\newpage

\bc
{\Large\bf Table II}\\[0.3cm]
\bt{|l|c|c|c|c|}\hline
{}~~~~~Reaction & Ref. & $P_L\;(GeV/c)$ &
$\sigma_{exp}(\mu b)$ & $\sigma_{theor}(\mu b)$\\ \hline
$pp\to D^+~X$ & \cite{32} & $400$ & $5.7\pm1.5$ & $4.16$ \\
&&&& \\

$pp\to D^-~X$ & \cite{32} & $400$ & $6.2\pm1.1$ & $5.54$ \\
&&&& \\

$pp\to D^0~X$ & \cite{32} & $400$ & $10.5\pm1.9$ & $7$ \\
&&&& \\

$pp\to\bar D^0~X$ & \cite{32} & $400$ & $7.9\pm1.5$ & $12.3$ \\
&&&& \\

$pp\to D^+/D^-~X$ & \cite{33} & $800$ & $33\pm7$ & $22.2$ \\
&&&& \\

$pp\to D^0/\bar D^0~X$ & \cite{33} & $800$ &
$26$\parbox{1cm}{$+21$\\ $-13$} & $45.4$ \\
&&&& \\

$pp\to D^+/D^-~X$ & \cite{34} & $800$ & $26\pm14$ & $22.2$ \\
&&&& \\

$pp\to D^0/\bar D^0~X$ & \cite{34} & $800$ & $22$\parbox{1cm}{$+4$\\
$-7$} & $45.4$ \\
&&&& \\

$pN\to D/\bar D~X$ & \cite{35} & $200$ & $1.5\pm0.7\pm0.1$ & $5.6$ \\
$x_F>0$ & & & & \\
&&&& \\

$\pi^-N\to D^+/D^-~X$ & \cite{35} & $200$ &
$1.7$\parbox{1cm}{$+0.4$\\ $-0.3$}$\pm0.1$ & $3.5$ \\
&&&& \\

$\pi^-N\to D^0/\bar D^0~X$ & \cite{35} & $200$ &
$3.3$\parbox{1cm}{$+0.5$\\ $-0.4$}$\pm0.3$ & $5.3$ \\
&&&& \\

$\pi^-N\to D^-/D^0~X$ & \cite{35} & $200$ &
$2.3$\parbox{1cm}{$+0.4$\\ $-0.3$}$\pm0.1$ & $4.7$ \\
&&&& \\

$\pi^-N\to D^+/\bar D^0~X$ & \cite{35} & $200$ &
$3.2$\parbox{1cm}{$+0.5$\\ $-0.4$}$\pm0.2$ & $4.2$ \\
&&&& \\

$\pi^-p\to D^+/D^-~X$ & \cite{36,37} & $360$ & $5.7\pm1.5$ & $7.76$ \\
&&&& \\

$\pi^-p\to D^0/\bar D^0~X$ & \cite{36,37} & $360$ & $10.1\pm2.2$ &
$11.0$\\ \hline
\et
\ec

\newpage

\bc
{\Large\bf Table III}\\[0.3cm]
\bt{|l|c|c|c|c|}\hline
{}~~~~~Reaction & Ref. & $P_L\;(GeV/c)$ & $\sigma_{exp}(\mu b)$ &
$\sigma_{theor}(\mu b)$\\ \hline

$pp\to D^{*+}/D^{*-}~X$ & \cite{32} & $400$ & $9.2\pm2.4$ & $7.14$ \\
&&&&\\

$pp\to D^{*0}/\bar D^{*0}~X$ & \cite{32} & $400$ & $5.8\pm2.7$ &
$8.8$ \\
&&&&\\

$\pi^-p\to D^{*+}/D^{*-}~X$ & \cite{38} & $360$ &
$5.0$\parbox{1cm}{$+2.3$\\ $-1.8$} & $5.0$ \\ &&&&\\

$\pi^-p\to D^{*0}/\bar D^{*0}~X$ & \cite{38} & $360$ & $7.3\pm2.9$ &
$4.5$ \\
&&&&\\

$\pi^-N\to D^{*+}/D^{*-}~X$ & \cite{35} & $200$ &
$2.4\pm0.4\pm0.2$ & $2.6$ \\
&&&&\\

$pp\to D^+_s/D^-_s~X$ & \cite{32} & $400$ & $<2.5$ & $2.8$ \\

{}~~~~~~$x_F>0$ & & & & \\ \hline
\et
\ec

\newpage
\section*{Table captions}

\noindent
\bt{p{1cm}p{15cm}}
I & comparison of the experimental cross sections of $\Lambda_c$
production in $pp$ and $\pi p$ interactions with results of our
calculations.\\[2mm]
II & The same as in Table I of $D$--mesons production.\\[2mm]
III & Experimental data and model calculations for $D^*$-- and
$D_s$--mesons production in $\pi p$ and $pp$ collisions.
\et

\newpage

\section*{Figure captions}

\noindent
\bt{p{1.5cm}p{14.5cm}}

Fig.1 & Diagrams corresponding to fragmentation of diquarks into
baryons.\\[2mm]

Fig.2 & Comparison QGSM calculations with experimental data on
$\Lambda_c$ spectra in: a) $\pi^-p\to\Lambda_cX$ at
$P_i=230\;GeV/c$ \cite{29}, b) $pp\to\Lambda_cX$ at $\sqrt
s=63\;GeV/c$ \cite{30,31}.\\[2mm]

Fig.3 & The $x_F$--distributions of different $D$--mesons in $pp$
collisions at $400\;GeV/c$ \cite{32}: a) $D^+$, b) $D^-$, c) $D^0$,
d) $\bar D^0$.\\[2mm]

Fig.4 & Inclusive distributions of all $D$--mesons in $pp$
interaction at a) $200\;GeV/c$ \cite{35}, b) $400\;GeV/c$ \cite{32},
c) $800\;GeV/c$ \cite{34}.\\[2mm]

Fig.5 & comparison of the model calculations with experimental data on
leading ($D^-/D^0$) and nonleading ($D^+/\bar D^0$) charmed mesons
in $\pi^-p$ interaction: a) leading, $200\;GeV/c$ \cite{35},
b) nonleading, $200\;GeV/c$ \cite{35}, c) leading, $360\;GeV/c$
\cite{37}, d) nonleading, $360\;GeV/c$ \cite{37}.\\[2mm]

Fig.6 & The $x_F$--dependence of $D^*$--meson in $\pi^-p$ interaction
at $360\;GeV/c$ \cite{38}: a) $D^{*+}/D^{*-}$--mesons, b)
$D^{*0}/\bar D^{*0}$--mesons.\\[2mm]

Fig.7 & Inclusive spectra of charmed baryons in $\Sigma^-p$ collision:
a) $\Lambda_c$, $\Xi^+_c$, $\Xi^0_c$, $\Omega_c$ at $340\;GeV/c$;
b) $\Xi^{*0}_c$, $\Xi^{*+}_c$, $\Omega^{*}_c$ at $340\;GeV/c$; c)
the same as in a) but for $600\;GeV/c$; d) the same as in b) but for
$600\;GeV/c$.\\[2mm]

Fig.8 & The same calculations as in Fig.7 for $\Xi^-$ beam.\\[2mm]

Fig.9 & Predictions for different $D$ and $D^{*}$--meson
production on $\Sigma^-$ beam: a) $D^+$, $D^-$, $D^0$, $\bar D^0$ at
$340\;GeV/c$; b) $D^{*+}$, $D^{*-}$, $D^{*0}$, $\bar D^{*0}$ at
$340\;GeV/c$; c) the same as in a) at $600\;GeV/c$; d) the same as in
b) at $600\;GeV/c$.\\[2mm]

Fig.10 & The same as in Fig.9 for $\Xi^-$ beam.\\[2mm]

Fig.11 & The $x_F$--dependence of $D_s$-- and $D^*_s$--mesons
production in $\Sigma^-p$ collisions: a) $D^+_s$ and $D^-_s$ at
$340\;GeV/c$, b) $D^{*+}_s$ and $D^{*-}_s$ at $340\;GeV/c$, c) as in
a) at $600\;GeV/c$, d) as in b) at $600\;GeV/c$.\\[2mm]

Fig.12 & The same as in Fig.11 for $\Xi^-$ beam.

\et

\end{document}